\documentclass[12pt]{article}
\usepackage{amsmath}
\usepackage{amsfonts}
\usepackage{amsthm}
\usepackage{amssymb}
\usepackage{graphicx}
\usepackage{enumerate}
\usepackage{natbib}
\usepackage{url} 
\usepackage{graphicx}
\usepackage{caption}
\usepackage{subcaption}
\usepackage{comment}
\usepackage{booktabs}
\usepackage[table,xcdraw]{xcolor}

\newcommand{\blind}{1}

\begin{document}

\def\spacingset#1{\renewcommand{\baselinestretch}%
{#1}\small\normalsize} \spacingset{1}


\if1\blind
{
  \title{\bf A Sensitivity Analysis Framework for Quantifying Confidence in Decisions in the Presence of Data Uncertainty}
  \author{Adway S. Wadekar \thanks{
    The authors gratefully acknowledge support from NSF SES 2217456.}\hspace{.2cm}\\
    Department of Statistics, University of Michigan\\
    and \\
    Jerome P. Reiter \\
    Department of Statistical Science, Duke University}
  \maketitle
} \fi

\if0\blind
{
  \bigskip
  \bigskip
  \bigskip
  \begin{center}
    {\LARGE\bf A Sensitivity Analysis Framework for Quantifying Confidence in Decisions in the Presence of Data Uncertainty}
\end{center}
  \medskip
} \fi

\bigskip
\begin{abstract}
Nearly all statistical analyses that inform policy-making are based on imperfect data. As examples, the data may suffer from measurement errors, missing values, sample selection bias, or record linkage errors.
    Analysts have to decide how to handle such data imperfections, e.g., analyze only the complete cases or impute values for the missing items via some posited model.  Their choices can influence estimates and hence, ultimately, policy decisions.  Thus, it is prudent for analysts to evaluate the sensitivity of estimates and policy decisions to the assumptions underlying their choices. 
    To facilitate this goal, we propose that analysts define metrics and visualizations that target the sensitivity of the ultimate decision to the assumptions underlying their approach to handling the data imperfections. Using these visualizations,  the analyst can assess their confidence in the policy decision under their chosen analysis. 
    We illustrate metrics and corresponding visualizations with two examples, namely considering possible 
    measurement error in the inputs of predictive models of presidential vote share and imputing missing values when evaluating the percentage of children exposed to high levels of lead. 
\end{abstract}

\noindent%
{\it Keywords:}  election, imputation, measurement, missing, nonignorable.
\vfill

\newpage

\section{Introduction}
There is no question that statistical analyses can inform policy-making.  Often, however, the data used for such analyses are messy and imperfect. Values of important variables may be missing or measured with error. Data files constructed by linking records across multiple databases may be prone to incorrect or incomplete linkages. The data themselves may be collected from a sample that is not representative of the full population.  

Data analysts must confront such imperfections as part of the analysis process.  This could involve explicitly accounting for data imperfections. For example, they may decide to impute missing values according to some (possibly missing not at random) model and subsequently analyze the completed datasets \citep{little2019statistical}. Or, they may decide to embed a measurement error model in their analysis \citep{fuller}.  Alternatively, they could  simply disregard the imperfections and treat the collected data ``as is'' in the analysis. For example, they could analyze only the complete cases or fix possibly erroneous values at their recorded values.  Even with such do-nothing choices, the validity of the resultant analyses still relies on implied assumptions, e.g., the data are missing or faulty completely at random \citep{reiter:raghu}. 

For some policy questions, analysts utilize data provided by a statistical agency. Frequently, such data are preprocessed by the agency, which corrects imperfections as best it can. This too represents a modeling choice: the analyst implicitly accepts the assumptions underpinning the agency's preprocessing steps \citep{meng:14}.



The analyst's choices in dealing with imperfections can influence their statistical inferences.  In turn, this ultimately can affect policy decisions that derive from those inferences. It is therefore  prudent for analysts to assess the sensitivity of statistical inferences, and arguably more importantly the ultimate decisions that stem from those inferences, to the reasonableness of the assumptions underpinning their approach for dealing with data imperfections. 

In this article, we present a framework to facilitate such sensitivity analyses.  The key idea is to define a metric that targets the sensitivity of the ultimate decision to the assumptions underlying the approach to handling the data imperfections. We emphasize metrics for decisions as opposed to statistical inferences, as ultimately it is the decision that matters most.  To illustrate, suppose a health agency seeks to estimate the percentage of people exposed to a communicable disease.  It may not matter if data imperfections cause the estimate to be off by, say,  5\%  when the true percentage is high; whether the estimate is 75\% or 80\%, the conclusion (decision) is that the disease has spread to a sizable fraction of the population.  On the other hand, in a different context, say political polling, an error of 5\% could lead to a grossly incorrect decision.  In these examples, the important metric is not the amount of error in an absolute sense---that is, purely statistical concepts like bias, mean-squared error, and confidence interval coverage---but the amount of error given the context of the decisions that will be made from the analysis. 

After defining a metric, analysts compute it under their proposed method for handling the data imperfections.  They also compute it under varying degrees of departure from the assumptions underlying their proposed method. In this way, analysts can assess how much of a departure is needed before the decision changes in practically meaningful ways.  In cases where plausible departures do not effect meaningful changes in decisions, analysts can reach their decisions with confidence.  Otherwise, analysts should have low confidence in the decisions. With this usage in mind, we refer to the framework as a confidence-in-decision (CID) analysis. 
    
The CID analysis framework has features in common with the types of sensitivity analyses often done for causal inference \citep{fogarty2023sensitivity} and for imputation of missing data. 
In causal inference, 
the goal of a sensitivity analysis typically is to determine how strong the effect of unmeasured confounding must be to nullify an apparent causal relationship \citep{rosenbaum, vanderweele2017sensitivity, ding2016sensitivity, li:reiter:scott, mathur2023m}. 
With missing data, the goal of a sensitivity analysis typically is to assess how much estimates change under different specifications of missing not at random mechanisms \citep{little2019statistical, linero:daniels, schifeling:reiter}.  \cite{liublinska2014sensitivity} propose tipping point analysis, which offers a visualization of how the $p$-value for a test of no treatment effect changes under increasingly serious departures from missing at random assumptions. Our work contributes to these approaches by turning the focus explicitly on the impacts of assumptions on ultimate decisions as opposed to statistical estimates alone.

The CID analysis framework also is related to statistical decision analysis \citep{berger2013statistical}.
Decision analyses 
map statistical estimates to a space of utilities via some function.  For example, \cite{suzuki} defines utilities that are a function of the size of the effect of some policy intervention, a probability distribution over possible effect sizes, the cost of implementing the intervention, and the cost of the event that the policy is intended to address. The goal of these analyses is typically to identify some optimal decision or strategy by maximizing the utility function.  Our framework  aims at revealing the sensitivity of ultimate decisions to the methods chosen to handle data imperfections as opposed to finding an optimal solution.

The remainder of this article is organized as follows.
In Section~\ref{sec:general}, we outline the CID framework.  We do in the context of potential measurement error in the input to a predictive model for presidential election outcomes.  
In Section~\ref{sec:illustrate}, we illustrate a  
a CID analysis with missing data. Here, the analyst considers costs of using the estimate from a missing at random (MAR) analysis when in fact the observed data are missing not at random (MNAR) to varying degrees. 
The context is the percentage of children in North Carolina exposed to high levels of lead. 
We conclude in Section~\ref{sec:discussion} with a discussion and avenues for future directions.  R code to reproduce the results is available at: \url{https://github.com/adway/CID-analysis}.

\section{The CID Analysis Framework Illustrated Via Measurement Error in Model Inputs} \label{sec:general}

We illustrate the general approach for CID analysis using the regarded Bread and Peace model of~\cite{hibbs2000bread}. In the Bread and Peace model, the vote share for an incumbent's party in a presidential election is regressed on a weighted-average growth of real income during the incumbent's presidential term. The model, while simple, offers remarkably reliable predictions~\citep{gelman2021regression}.  However, the input to the model is in fact an estimate and a potentially noisy one at that; estimates of economic indicators generally have been shown to be subject to measurement errors   
\citep{angel2019did, moore2000income}. 
As with any regression model, the accuracy of the prediction depends on the accuracy of the input. 
Thus, we are confronted with the CID analysis questions: 
what happens if the measurement of the income variable is erroneous, and how inaccurate must the measurement be in order to change our decision in predicting a future presidential election based on this model? 

\begin{figure}[t]
    \centering
    \includegraphics[width=0.7\textwidth]{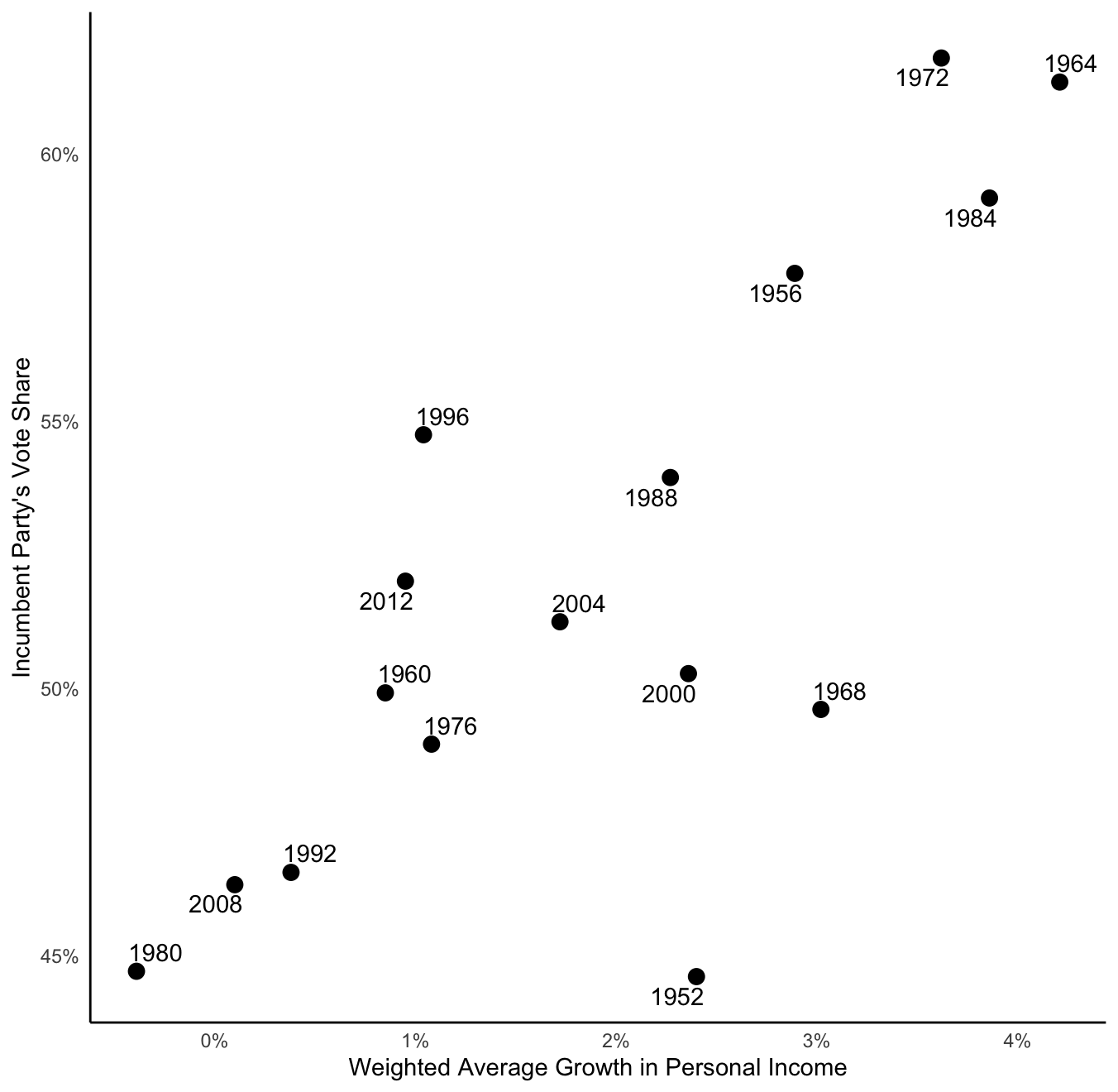} 
    \caption{Data used in Bread and Peace model for predicting incumbent vote share from economic growth.}
    \label{fig:hibbs-explore}
\end{figure}

We estimate the Bread and Peace model using the data provided in~\cite{gelman2021regression}, displayed in Figure~\ref{fig:hibbs-explore}. The data cover every presidential election from 1952 to 2012. The data include the weighted average growth rate in real income over the previous presidential term 
along with the vote share won by the incumbent's party. The estimated model is
\begin{align}
	\text{Vote Share}_i = 46.248 + 3.061 \times \text{Weighted Average Growth}_i + e_i, \quad e_i \sim \mathcal{N}(0, \hat{\sigma}^2 = 14.16), \label{BnP}
\end{align}
where $i$ corresponds to an index for an election. 

Using \eqref{BnP}, suppose that we seek 
to predict vote share in the 2024 presidential election from weighted average growth in real personal income throughout the previous four years, which we estimate with data from the Saint Louis Federal Reserve. In the 2024 election cycle, this input corresponds to $x = -0.728\%.$  Thus, the predicted vote share for the incumbent's party (Democrat) in 2024 is 44.0\% with a 95\% prediction interval of (39.6\%, 48.4\%).  
The prediction interval clearly suggests that the challenger is favored to win (which indeed occurred). This is the decision reached from the analysis.

The prediction that the challenger will win presumes no measurement error in $x$.  To assess the sensitivity of the decision to potential errors in the input variable, we perform a CID analysis.  We first require a CID metric, which we now develop.

When predicting elections, 
analysts can make one of three types of decisions: candidate A will win, candidate B will win, or it is not possible to predict a clear winner. 
We construct a decision rule as a map $\varphi(\ell, u)$ 
that takes the prediction interval $(\ell, u)$ for the vote share ($\times 100$), where $\ell$ and $u$ are the lower and upper bounds of the interval, respectively,  to a decision as follows,
\begin{align}
	\varphi(\ell, u) =
\begin{cases} 
    \text{incumbent wins}, & \text{if } 50 < \ell < u \\  
    \text{unclear who wins}, & \text{if } \ell < 50 < u \\  
    \text{challenger wins}, & \text{if } \ell < u < 50. 
\end{cases} \label{decisionvote} 
\end{align}
The rule in \eqref{decisionvote} decides the incumbent wins the election when the lower and upper bounds of the prediction interval 
are both above 50\%, decides it is unclear who wins when the prediction interval includes 50\%, and decides the challenger wins when the bounds 
are both below 50\%. 

 We are interested in the impact of measurement error in the input to \eqref{BnP} on our confidence in the decision in \eqref{decisionvote}. To assess this impact, we perturb $x$ 
 by a parameter $t$   
 corresponding to additive measurement error. For example, a value of $t = 0.5$ corresponds to an analyst suspecting that the true value of the growth in personal income is 0.5 percentage points higher than the value $x$ used to make the prediction. The value $t=0$, which for convenience we label as $t_0$, corresponds to no measurement error.  Absent specific information about the nature of the measurement error, we use $t_0$ for the baseline prediction from the Bread and Peace model.  We then compute predicted vote share and associated prediction intervals for several  plausible values of the true growth, $x + t$.  \textcolor{black}{For example, with $t=0.5$, the predicted vote share is $46.248+3.061(-0.728+0.5)$.} We refer to this perturbation and computation step as ``turning the knob'' on the amount of measurement error.  More extreme values of $t$ correspond to more extreme measurement error. For each of these perturbed inputs, we compute the CID metric.

To construct the CID metric for this analysis, we first develop a function that indicates whether or not the analyst's decision about the election is the same.
For any $t$ under consideration, let $D_{t}=1$ when $\varphi(\ell, u)$ in \eqref{decisionvote} remains the same at $t$ as it does at $t_0,$ and let $D_{t}=0$ otherwise.  The value of $D(t)$ depends on the decision made by the analyst with their proposed approach to handling potential  measurement error (in this case, disregarding it).  In our analysis, because the reference analysis predicts the challenger to win and \eqref{BnP} has a negative slope, $\varphi(\ell, u)$  remains unchanged if in fact the observed value of weighted average growth in income overestimates the true value (i.e., $t<0$), regardless of the amount of overestimation. On the other hand, when $x$ in fact underestimates the true value of the predictor (i.e., $t>0$), $\varphi(\ell, u)$  continues to favor the challenger for some values of $t$ until eventually switching to uncertainty (when the prediction interval includes 0.5) or favoring the incumbent. 


One can use $D_t$ as a CID metric. In some cases, however, analysts may want a more nuanced metric that reflects changes in the prediction intervals as one turns the knob for $t$.  With this in mind, we construct a CID measure that assigns greater confidence in the decision when the prediction interval for some $t \neq t_0$ overlaps substantially with the prediction interval for $t_0$, and it assigns lesser confidence in the decision when the intervals based on $t$ and $t_0$ have little or no overlap.  To encode this, we 
incorporate the confidence interval overlap metric of~\cite{karr2006framework}.
For any $t \neq t_0$, let $(L_{t}, U_{t})$ and $(L_{t_0}, U_{t_0})$ be the prediction intervals using $t$ and $t_0$, respectively, in \eqref{BnP}. Let $(L_{\mathrm{over}}, U_{\mathrm{over}})$ represent their intersection, defined respectively as the minimum and maximum values in the set, $\{b: b \geq L_{t_0}, b \geq L_{t}, b \leq U_{t_0}, b \leq U_{t} \}$. When the set is empty, we let  $L_{\mathrm{over}} =  U_{\mathrm{over}}$. The overlap in the prediction intervals derived from the analysis with $t_0$ and the analysis with $t$ is 
\begin{align}
	J_{t} = \frac{1}{2}\left[\frac{U_{\mathrm{over}} - L_{\mathrm{over}}}{U_{t_0} - L_{t_0}} + \frac{U_{\mathrm{over}} - L_{\mathrm{over}}}{U_{t} - L_{t}}\right].\label{cioverlap}
\end{align}	
The value of $J_t$ takes a maximum of 1 when the two intervals have the same lower and upper bounds, and a minimum of zero when the intervals have no overlapping region.  Confidence interval overlap measures were developed originally for statistical disclosure limitation, where one interval is based on the confidential data and the other is based on perturbed or synthetic data.  Large values of the overlap measures, say above 0.5, are taken as evidence that the disclosure-protected and confidential data offer similar inferences for the estimand under consideration.

We combine the confidence interval overlap with the indicator of decision change.  The result is the CID metric,
\begin{align}
		\mathrm{CID}(t) = D_t\left(1 + J_t \right).\label{defn:cid-general}
	\end{align}

When $\varphi(\ell, u)$ changes for a particular $t$, $D_t=0=\mathrm{CID}(t)$ and the analyst should have no confidence in their decision at that level of measurement error. When $\varphi(\ell, u)$  does not change for a particular $t$, $D_t=1$ and $1 \leq \mathrm{CID}(t) \leq 2$. Naturally, $\mathrm{CID}(t)$ reaches its highest value of 2 when $t=t_0$, i.e., there is no measurement error in $x$. When the measurement error affects the prediction intervals but not $\varphi(\ell, u)$, $\mathrm{CID}(t)$ decreases from 2 to a minimum of 1 as the intervals decrease in overlap, representing a 
cost in decision confidence.





\begin{figure}[t]
    \centering
    \includegraphics[width=0.8\textwidth]{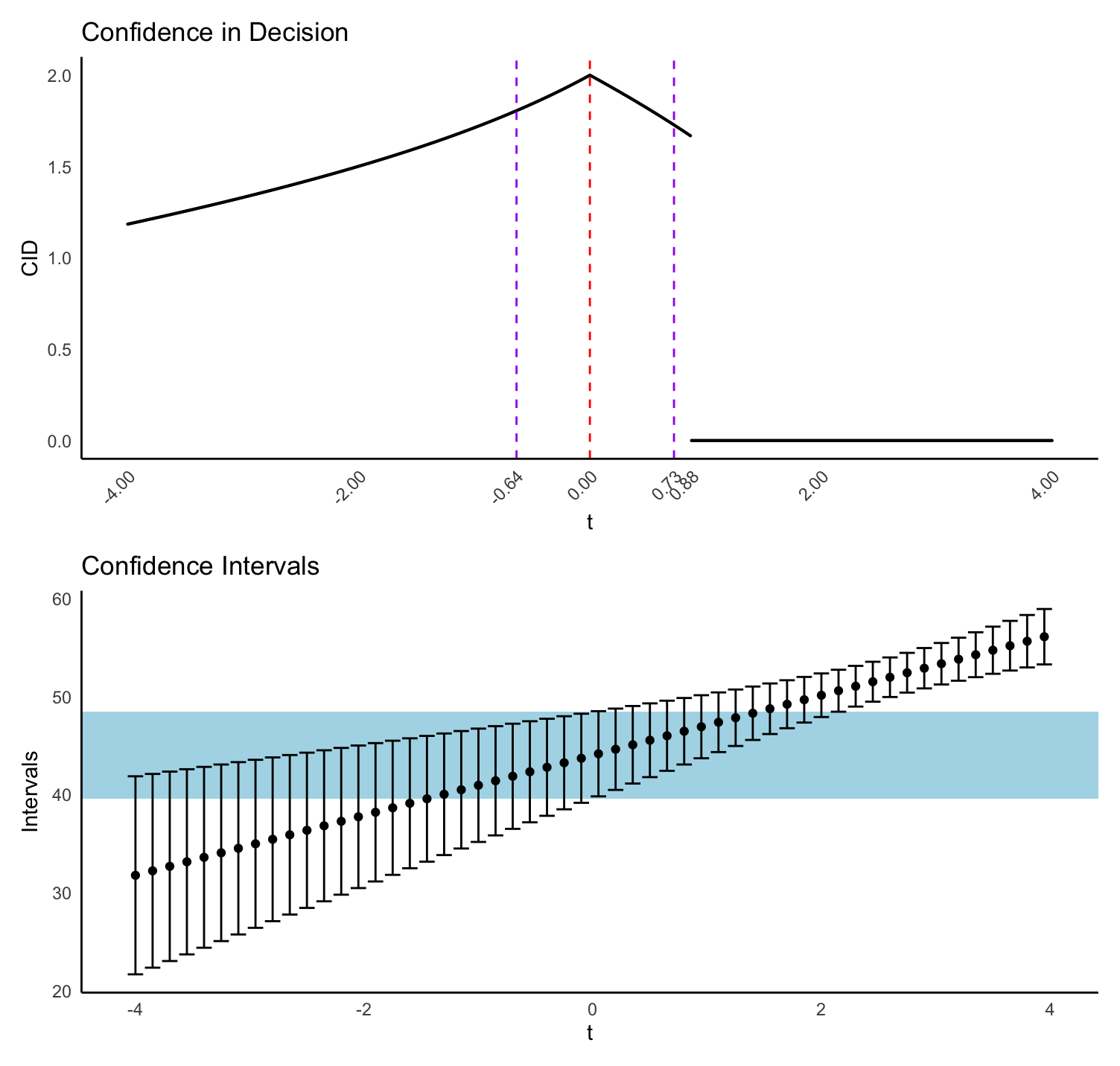} 
    \caption{Confidence in decision (CID) analysis for the 2024 presidential election prediction using the Bread and Peace model.  Top panel displays the change in the CID metric for various amounts of measurement error. Red line indicates the value of the input with no measurement error.  Purple lines indicate the boundaries of a region of plausible  measurement error as determined by annual variability in the recorded income measures over the preceding four year period.
    Bottom panel displays the  prediction intervals for the values of $t$ used in the top panel. Blue bar is the prediction interval for the reference analysis, which presumes the input has no measurement error.}
    \label{fig:cid-income}
\end{figure}

We now use $\mathrm{CID}(t)$ to assess our confidence in the 2024 presidential election prediction from the Bread and Peace model.
Figure~\ref{fig:cid-income} displays the values of the metric for $-4 \leq t \leq 4$, assuming no measurement error as the reference analysis. 
This corresponds to a possible  overestimate or underestimate in the real weighted average income growth of up to 4 percentage points. As evident in the top panel,  
 once the measurement error $t > 0.88$, so that the wage growth is truly at least 0.172, the decision  changes, first to unclear winner at $t= 0.88$ and then to the incumbent at $t=2.62$. For  $-4 \leq t \leq 0.88$, the ultimate decision to predict the challenger as the winner does not change. That is, the CID metric does not drop to zero over this range of $t$.  The prediction interval based on $t_0$, i.e., no measurement error, overlaps substantially with the prediction intervals for many values of $t$; for example, $J_t \geq 0.5$ for $-2 \leq t \leq 1.2$.  Note that when  $0.88\leq t \leq 1.2$, we have $J_t>0.5$ but the decision itself changes to an unclear winner, making $D_t=\mathrm{CID}(t)=0$ for $t$ in this range. 
Analysts can use Figure \ref{fig:cid-income} to visualize how quickly the confidence in decision changes with increasing measurement error. To utilize this figure most effectively, analysts need to 
determine what range of measurement errors is plausible.  In other words, how far is too far to turn the knob for $t$?
Ultimately, specifying plausible ranges of $t$ for a specific CID analysis should be based on domain expertise. 
For this problem, we define a plausible range of error by considering the widest range of growth or loss in the four year window for

\clearpage

\noindent which the weighted average was computed. In the case of the 2024 election, using election day in 2019 as a starting point, the largest loss was 1.363\% and the gain never crosses 0\%.  This corresponds to a plausible region for $t$ of  $(-0.635, 0.728)$.  As evident in Figure~\ref{fig:cid-income}, the CID metric does not change substantially within this plausible range of measurement error. As a result, the analyst can conclude with reasonable confidence that their decision with the Bread and Peace model is insensitive to plausible measurement error in $x$.

\textcolor{black}{Analysts can use  CID analysis with other ways of accounting for potential measurement error. As one example, to make a prediction interval for the 2024 vote share, analysts can specify and average over a distribution of the measurement error for $x$, such as an additive measurement error model where $t \sim N(0, \tau^2)$ for some selected variance $\tau^2.$  By averaging over this distribution, the analyst can obtain a single prediction interval with upper and lower limits that can be inputs to the CID metric. Analysts could repeat the analysis using different measurement error distributions, e.g., via different values of $\tau^2$, in which case $\tau^2$ becomes the knob to enable a CID analysis.}

We close this section by summarizing the general approach to CID analysis.
\begin{enumerate}
	\item {\bf Specify a rule that maps the estimator of interest to a decision.} The analyst applies this decision rule to the estimate from their proposed approach to handling the data imperfections to arrive at a reference decision. For the Bread and Peace model, this corresponds to computing the prediction interval for the growth value taken as is, i.e., presuming no measurement error, and mapping the prediction interval to a decision about the election outcome. 
	\item {\bf Compute estimates under plausible departures from the assumptions underlying the reference analysis.} In many cases, this involves specifying a sensitivity parameter that can be varied to represent departures from the assumptions used in the reference analysis.  For the Bread and Peace model, this corresponds to computing prediction intervals for posited true growth values, which are determined by adding $t$ to the original input $x$. 
	\item  {\bf Construct a CID metric.} The CID metric should reflect what the analyst needs to assess their confidence in the results under  departures from the assumptions used in the reference analysis.  
    For the Bread and Peace model, the CID metric reflects both the potential for a decision change and the disparity in prediction intervals constructed under an assumption of no measurement error versus an assumption of measurement error of a particular level $t$.
	\item  {\bf Assess confidence in the  reference decision using the values of the CID metric.} Visualizations can help the analyst determine when the CID metric attains values that cause a lack of confidence. For these values, the analyst can assess whether the departures from the assumptions are plausible.  
    For the Bread and Peace model, this corresponds to specifying a range for the measurement error $t$ that accords with the variability in the measured growth over the previous four year period.
\end{enumerate}

As with all sensitivity analyses, a critical question is how to assess the plausibility of various departures from the baseline assumptions.  For example, how much unmeasured confounding is plausible in a causal inference sensitivity analysis, or how strongly does a missingness mechanism depart from MAR assumptions for a missing data sensitivity analysis? It may be helpful to calibrate plausible departures by comparing to other known effects. For example, in sensitivity analysis in causal inference, one can compare the posited effects of unmeasured confounding to the strength of associations between the outcome and selected covariates; in missing data sensitivity analysis, one can compare posited differences in the distributions of missing and observed values to distributional differences across demographic groups. 

\section{Example of CID Analysis with Missing Data} \label{sec:illustrate}

We now illustrate a CID analysis where the data imperfection is missing data.  Typically, analysts presume that values are MAR. This may be partly because standard statistical methods and software are designed for missingness at random. Regardless, the analyst may be concerned that a decision reached under MAR assumptions could be incorrect if the data are MNAR and hence seek to assess their confidence in the decision under departures from a default MAR modeling approach. We illustrate such analyses using data on lead exposure among children in North Carolina, as we now describe.




Exposure to high levels of lead can have adverse effects on children's cognitive development~\citep{bellinger1992low, miranda2009environmental, bravo2022racial}. Thus, public health decision-makers may choose to intervene when the percentage of children exposed to high levels of lead is deemed unacceptable.
To determine these percentages, one approach is to utilize large administrative health databases, which may contain tens of thousands of tested children. However, there can be selection mechanisms into these databases; for example, children who are suspected to have high levels of lead exposure may be more likely to be tested for that exposure. Thus, an analyst computing lead exposure rates from administrative data should evaluate their confidence in their ultimate decision about the percentage of children at unacceptably high levels, as well as any resulting policy intervention. 



\begin{table}[t]
    \centering
    \begin{tabular}{c c c c c c c c c c c}
        \toprule
        Lead Level & 1 & 2 & 3 & 4 & 5 & 6 & 7 & 8 & 9 & 10 \\
        \midrule
        Probability & 0.22 & 0.31 & 0.22 & 0.13 & 0.04 & 0.03 & 0.02 & 0.01 & 0.01 & 0.01 \\
        \bottomrule
    \end{tabular}
    \caption{Empirical distribution of lead exposures (micrograms per deciliter) among the children  in the North Carolina lead data. Smaller values indicate lesser amounts of lead exposure. Estimates taken from Feldman et al. (2024). The blood lead measurements in the data are available only as integers; we do not have access to values with greater precision.}
    \label{tab:lead-levels}
\end{table}

For our analysis, we use administrative data from the state of North Carolina for children born between 2003 and 2005.  These data were used by \cite{feldman2024gaussian}, among others, to assess correlates of lead exposure and educational outcomes. The data comprise approximately $n=110,000$ children in the desired age range who have recorded, integer-valued blood lead levels.  Table~\ref{tab:lead-levels} displays the empirical percentages.  According to the \cite{children}, approximately $N=400,000$ children were born in North Carolina between 2003 and 2005.  Hence, considering the $n$ children as part of the population, we are missing lead measurements for approximately $290,000$ children. Here, we presume no other information is available on these children. This is actually the case for us as we do not have access to the record-level data, which are confidential. 


One strategy for handling the missing data is to use multiple imputation \citep{rubin:1987}.  For $i=1, \dots, N$, let $y_i$ be the lead exposure measurement for the $i$th child in North Carolina in the age range of interest.
For $i=1, \dots, N$, let $r_i=0$ when $y_i$ is observed and $r_i=1$ when $y_i$ is not observed. Thus, the $n$ recorded lead measurements comprise $\{y_i: r_i=0\}$.  
Because the child-level data are unavailable to us, we simulate observed lead values using the empirical distribution in Table~\ref{tab:lead-levels}.  For $k=1, \dots, 10$, let $n_k$ be the number of individuals in the simulated data to be at lead level $k$.

As the strategy for handling the missing data, we presume the analyst follows typical practice and applies a missing (completely) at random model for multiple imputation.  
We assume that lead levels follow a multinomial distribution with support on the integers from 1 to 10. We use uniform prior distributions on the multinomial probabilities.  Thus, to make any single completed dataset comprising the $n$ observed and $N-n$ imputed lead values, under MAR we first draw a value of the $10$-dimensional multinomial probability vector $\mathbf{p} \sim \mathrm{Dirichlet}(1 + n_1, \dots, 1 + n_{10})$.  Given the sampled value of $\mathbf{p}$, we draw a value $y_i \sim \mathrm{Multinomial}(\mathbf{p})$ for all units with $r_i=1.$  We repeat this procedure independently $M=5$ times to create $M$ completed populations, $D^{(1)}, \dots, D^{(M)}$.  The analyst can use the combining rules from \cite{rubin:1987} to make inferences about any estimands, including ones intended to inform a policy-maker's decision to intervene or not.


We now proceed with a CID analysis of the decision that the percentage of children in the relevant age group is high enough to warrant intervention.

\subsection{Step 1: Specify decision rule}

To construct an illustrative decision rule, we presume that a governmental agency provides intervention services when 20\% or more of the children in the state have lead exposure levels beyond 3.  This decision rule is motivated by targets for lead exposure rates in the Healthy People 2030 guidelines~\citep{lead}.  In practice, of course, public policy-makers could select the level or percentage that they deem most suitable to trigger interventions for the specific matter at hand.  
Our decision rule is a map $\varphi(\hat{\theta})$
that takes the estimated percentage $\hat{\theta}$ of children  with lead level above 3 to a decision as follows.
\begin{align}
	\varphi(\hat{\theta}) = \begin{cases} 
    \text{intervene}, & \text{if } \hat{\theta} > 0.20 \\  
    \text{don't intervene}, & \text{if } \hat{\theta} \leq 0.20. \label{eq:decthetahat}
\end{cases}
\end{align}

In each $D^{(m)}$ where $m=1, \dots, M$, the analyst computes the percentage of children whose lead level exceeds 3, $\hat{\theta}^{(m)} = \sum_{i=1}^N I(y_i^{(m)} > 3)/N$, where $I(a) =1$ when the operand $a$ is true and $I(a)=0$ otherwise. The multiple imputation point estimate is $\hat{\theta} = \sum_{m = 1}^M \hat{\theta}^{(m)}/M.$   In our application, $\hat{\theta}=0.25.$   Thus, the decision based on the MAR analysis is to intervene; the fraction of children with lead exposure levels above 3 appears to be quite high.  We note that the analyst also could compute a multiple imputation confidence interval for the population percentage; however, since each completed dataset represents a census, the multiple imputation variance is small as is the width of the interval.  It is sufficient in this case to base decisions on the point estimate.


\subsection{Step 2: Estimate under departures from missing at random}

Given the potential selection bias in the observed data, how confident should the analyst be in the decision that the percentage of children in the state who have lead levels above 3 exceeds 20\%?  To assess the confidence in this decision, we require a knob that allows us to encode departures from the MAR assumption underlying the observed data results. To devise this knob, we rely on the pattern mixture model approach to nonignorable missing data \citep{little2019statistical}, in which we explicitly specify some distribution for the missing lead values that differs from the observed data distribution.  

Specifically, we work with a log-odds representation of the probabilities in the multinomial distribution.  For $k=2, \dots, 10$, let  
$\log(p_k/p_1) = \beta_k$ 
so that 
\begin{eqnarray}
p_k &=& \exp(\beta_k)/(1+\exp(\sum_{k=2}^{10}\beta_k))\\
p_1 &=& 1/(1+\exp(\sum_{k=2}^{10}\beta_k)).
\end{eqnarray}
We sample values of $(\beta_2, \dots, \beta_{10})$ by sampling $\mathbf{p}$
from the Dirichlet distribution based on the observed data, and convert the draw to the log odds scale using the lead level of 1 as the baseline category. Define the sampled value as  
\begin{align}
	\boldsymbol{\beta}_0 = (0, \beta_2, \dots, \beta_{10}) = \left(0, \log(p_{2}/p_{1}), \dots, \log(p_{10}/p_{1})  \right).
	\end{align} 
We represent a MNAR mechanism by adding some  vector $\boldsymbol{\alpha}_t = (\alpha_{1}, \dots, \alpha_{10}) \neq \bf{0}$ to $\boldsymbol{\beta}_0$; we examine two choices for $\boldsymbol{\alpha}_t$ below.  Let   $\boldsymbol{\beta}_t = \boldsymbol{\beta}_0 + \boldsymbol{\alpha}_t$.  We use $\boldsymbol{\beta}_t$ to define the probabilities for the missing data imputation. That is, we impute each $y_i$ with $r_i=1$ using the multinomial distribution with probabilities 
	\begin{align}
		\boldsymbol{p}_t = \left(\frac{\exp(\alpha_{1})}{\exp(\alpha_1) + \sum_{k = 2}^{10} \exp(\beta_{k}+\alpha_k)}, \dots, \frac{\exp(\beta_{10}+\alpha_{10})}{\exp(\alpha_1) + \sum_{k = 2}^{10} \exp(\beta_{k}+\alpha_k)}\right).
	\end{align}
We repeat this process $M$ times to create $D^{(1)}, \dots, D^{(M)}$, and estimate  $\hat{\theta}^{(m)}$ in each.  The resulting average of the completed-data point estimates is $\hat{\theta}_t$, which derives from the multiple imputation combining rules described previously. 

One can set $\boldsymbol{\alpha}_t$ in literally an infinite number of ways; there is no evidence in the data alone to inform the choice.  Rather than try to determine the ``correct'' form of $\boldsymbol{\alpha}_t$, we instead suggest examining readily interpretable forms that facilitate understanding \citep{gelman:hennig, suzuki}, which in this case means the sensitivity of decisions to increasingly severe departures from MAR. Here we propose two such forms.

The first is a simple ``accordion-style'' missingness mechanism. We let $\alpha_{k} = t$ for $k < 4$ and $\alpha_{k} = 0$ for $k \geq 4$. This mechanism is motivated by the decision metric, in that what determines the decision is the fraction of children with lead levels above 3.  Large values of $t$ imply that the children without measurements of lead levels are more likely to have low levels, which is in line with the expected selection bias. 
The second is a more nuanced missingness mechanism that relies on a parametric form.   
    In particular, we examine $\boldsymbol{\alpha}_t = (t, 0.9t, 0.8t, 0.6t, 0.4t, 0, 0, 0, -0.2t, -0.25t).$
    The precise form of the chosen function is somewhat arbitrary, but this form encodes a mechanism for which positive and large values of $t$ correspond to 
    increased probabilities that children not tested have the lowest lead levels. 

\subsection{Step 3: Construct CID metric} 


We next construct the CID metric.
Let $D_t=1$ when the decision to intervene or not as determined by   \eqref{eq:decthetahat} is the same under the MAR  mechanism ($t=t_0=0$) and under the MNAR mechanism encoded using $t \neq 0$.
We use a $\mathrm{CID}(t)$ that depends on the value of $\hat{\theta}$ in the reference analysis---in our case, the estimate assuming MAR---as we  now describe.  

\textcolor{black}{First consider when   $\hat{\theta} \geq 0.20.$  In this case, the decision is to intervene. We presume the decision-makers intend to dedicate sufficient resources to the intervention to reduce the percentage of children with high lead exposure $\theta$ to the target of 20\%.  If the true $\theta$ happens to be less than $\hat{\theta}$, i.e., $\hat{\theta}$ is an overestimate, the decision-makers would  spend resources on the intervention unnecessarily;
this should be considered a negative in the CID metric. 
On the other hand, if $\theta \geq \hat{\theta}$, the decision makers would not meet the target; this too should be considered a negative in the CID metric. We encode these considerations using 
	\begin{align}
		\mathrm{CID}(t) = I(\hat{\theta}>.20)\left(1 -  \frac{I(\hat{\theta} \geq \hat{\theta}_t)\min(\hat{\theta} - 0.20, \hat{\theta} - \hat{\theta}_t)a + I(\hat{\theta}< \hat{\theta}_t)(\hat{\theta}_t - \hat{\theta})b}{C}\right),
        \label{costunder}
	\end{align}  
where $I(q) =1$ when the condition $q$ is true and $I(q)=0$ otherwise. Here, $a$ and $b$ are cost parameters set by the analyst and decision-makers.  Specifically, $a$ represents the cost of spending resources unnecessarily on an intervention, which is triggered by the overestimation of $\theta$.  And, $b$ represents the cost of not intervening to reduce the fraction of children with high lead exposure to the target level of 20\%.  In our analysis, we interpret $a$ and $b$ as costs per 1\% of children estimated to have high lead exposure. Additionally, $C$ is the maximum cost of inaccurate estimation. In \eqref{costunder}, we  set  
$C = \max((\hat{\theta} - .20)a,  (\theta_{\mathrm{WC}} – \max(\hat{\theta}, 0.20))b)$, where $\theta_{\mathrm{WC}}$ is the worst case percentage of children whose levels exceed 3 if every single child with $r_i=1$ has a lead level above 3.
Thus, the $\mathrm{CID}(t)=1$ when $\theta = \hat{\theta}$ and reduces to zero for the $\theta$ with the largest cost possible.}

\textcolor{black}{To understand this CID metric further, consider the value of \eqref{costunder} when $\hat{\theta}=0.25$ and the true percentage of children with high lead exposure is one of $\hat{\theta}_t \in \{0.15, 0.22, 0.27$\}.  For $\hat{\theta}_t=0.15$, the decision should be not to intervene, and the value of \eqref{costunder} reflects the $5a$ cost of that unnecessary intervention. For $\hat{\theta}_t=0.22$, the decision to intervene is right, but an additional $3a$ in cost was incurred unnecessarily. For $\hat{\theta}_t= 0.27$, $D_t=1$ but an additional 2\% of children have high lead exposure than estimated, leading to $2b$ in costs. These additional costs would be scaled by $C$ to facilitate interpretations and visualizations.} 


\textcolor{black}{
We also need to specify the CID metric when $\hat{\theta} <  0.20$. In this case, the decision under the reference MAR analysis is not to intervene.   
    The decision-maker incurs no cost for any value of $\theta$ that does not change the decision, i.e., for which $D_t=1$. However, when $\theta>.20$, $D_t=0$ and the decision maker should have dedicated sufficient resources for intervention to meet the target.  
We encode this by setting
    \begin{align}
		\mathrm{CID}(t) = I(\hat{\theta}<0.2)\left(1 - \frac{(1-D_t)(\hat{\theta}_t-0.20)b}{C}
        \right).
        \label{costover}
	\end{align}
    To illustrate, suppose that $\hat{\theta}=0.18$. For $\hat{\theta}_t = 0.19$, the CID metric in \eqref{costover} encodes no loss, since not intervening is still the right decision to stay under the target. For   $\hat{\theta}_t=.27,$ the decision-maker should have intervened, and the cost of missing 7\% of kids above the 20\% target is $7b$.  This $7b$ cost is scaled by the maximum possible cost of $C$.}

{In our analysis,  we set $a=b$, equating the cost of failing to provide enough resources for 1\% of children to not spending $a$ in resources. Since $\hat{\theta}= .25>.20$, we use \eqref{costunder} with $C=.54$
as ${\theta}_{WC} = 0.79$. 
Of course, this value of ${\theta}_{WC}$ is completely unrealistic; however, $C$ solely serves as a scaling factor for purposes of visualization.   
Analysts could set $(a,b)$ otherwise based on policy evaluations \citep{leadcosts}. More generally, the selection of $(a,b)$ is specific to the context at hand; for example, they could be based on valuations of the trade off in risks to child health and financial expenditures.}

\subsection{Step 4: Assess confidence in decision}
    
\begin{figure}[t]
    \centering
    \begin{subfigure}[b]{0.60\textwidth}
        \centering
        \includegraphics[width=\linewidth]{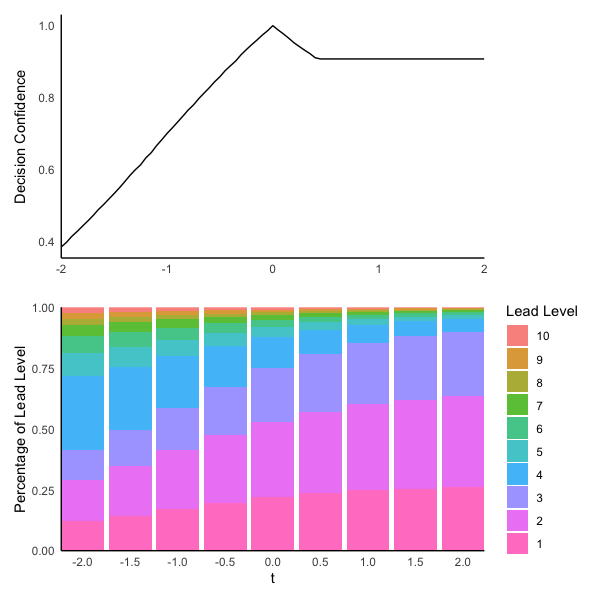}
        \caption{Accordion-style mechanism}
        \label{fig:param2}
    \end{subfigure}
    \hfill
    \begin{subfigure}[b]{0.60\textwidth}
        \centering
        \includegraphics[width=\linewidth]{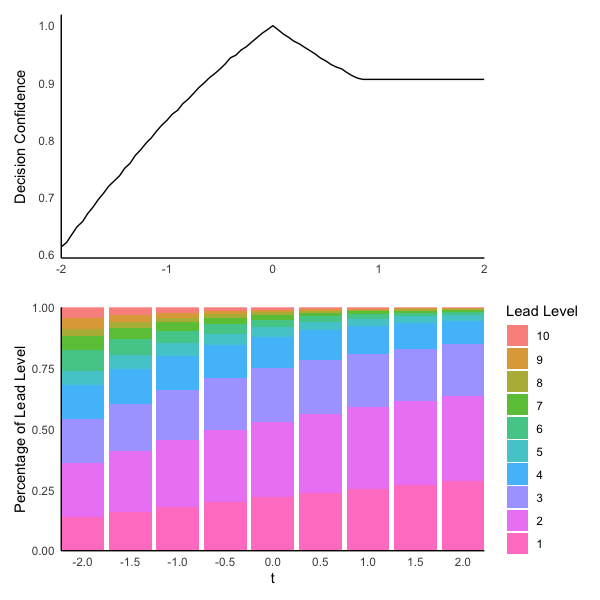}
        \caption{Parametric mechanism}
        \label{fig:param1}
    \end{subfigure}
    \caption{CID analyses of the North Carolina lead data under two posited MNAR mechanisms. The accordion-style mechanism uses $\boldsymbol{\alpha}_t = (t, t, t, 0, 0, 0, 0, 0, 0, 0).$ The parametric mechanism uses $\boldsymbol{\alpha}_t = (t, 0.9t, 0.8t, 0.6t, 0.4t, 0, 0, 0, -0.2t, -0.25t)$.}
    \label{fig:cid-lead}
\end{figure}

	Having constructed the CID metric, we can investigate the analyst's confidence in their reference decision (to intervene) when data are  MNAR.  We do so over ranges of $t$ for the two missingness mechanisms introduced previously. 
    Figure~\ref{fig:param2} and Figure~\ref{fig:param1} display the results. 
    These figures also display  for five values of $t$ the multiple imputation point estimates of the empirical frequencies of the lead levels among the $N$ children. 
    The decision changes around  $t=0.4$ under the accordion-style missingness mechanism and around $t=0.8$ under the parametric-style missingness mechanism. As shown in Table~\ref{table:change-point-lead}, 
    for the accordion-style mechanism, a value of $t=0.5$ generates an estimate of 19\% of children  
    having lead levels 
    above three, which is below the decision-rule  threshold. For the parametric-style missingness mechanism, a value of $t=1$ results in a similar 19\% figure, 
    though the distribution is somewhat more concentrated at levels of 2 and 3. {The cost of this overestimation is visualized as $.05/.54 \approx .10$, which is incurred for all values of $t$ that cause $D_t=0$.  The figures also reveal large costs that accrue quickly for $t<0$, which would occur if, for some reason, in actuality more than 25\% of children have lead levels exceeding 3.} 

    \begin{table}[t]
    \centering
    \begin{tabular}{l c c c c c c c c c c c}
        \toprule
        Missingness Mechanism & 1 & 2 & 3 & 4 & 5 & 6 & 7 & 8 & 9 & 10 \\
        \midrule
        Accordion ($t = 0.5$) & 0.24 & 0.34 & 0.24 & 0.10 & 0.03 & 0.02 & 0.02 & 0.007 & 0.008 & 0.008 \\
        Parametric ($t = 1$) & 0.26 & 0.33 & 0.22 & 0.11 & 0.03 & 0.02 & 0.01 & 0.006 & 0.006 & 0.005 \\ 
        \bottomrule
    \end{tabular}
    \caption{Empirical distribution of lead exposure for the $N$ children after multiple imputation  based on the accordion and parametric missingness mechanisms for values of $t$ close to where the decision changes from intervention to no intervention.}
    \label{table:change-point-lead}
\end{table}
    
    To evaluate the plausibility of such departures from MAR, we turn to reference lead levels among the population of children in the United States published by the Centers for Disease Control (CDC), as reported by \cite{feldman2024gaussian}. In particular, the CDC estimates that around half of U.S.\ children have levels less than or equal to 1, around 75\% have levels less than or equal to 2, and around 90\% have levels less than or equal to 3. If the population lead levels in North Carolina in fact are similar to  these U.S.\   percentages, we would have values of $t$ that represent substantial selection bias.  For example, with the parametric missingness mechanism, a value of $t\geq 10$ results in completed-data estimates

    %

\clearpage  
   
\noindent near the CDC percentages. However, as evident in Table \ref{table:change-point-lead}, the CID analysis indicates that the decision to intervene changes at much smaller values of $t$ than $t=10$.
Further, the reference decision about intervention could incur relatively large costs in the CID metric. For example, if we set $t=2$, we have a true percentage of children with lead levels exceeding 3 of 15\%.  Thus, by using the MAR estimate of 25\%, if $t=2$ we would be deciding to intervene unnecessarily, incurring 
{around 10\% of the maximum possible cost of an incorrect decision based on $\hat{\theta}=0.25$.}
   Translating this to a more practicable scale, if the state would spend, say, \$10 million per percentage point above the true population percentage,
   they would be spending an additional \$50 million when the reality might not dictate such intervention. Because of the decision change at values of $t$ that seem plausible, and because we would incur a substantial cost at those values, we recommend that the analyst should not have confidence in the decision to intervene under the MAR analysis.

\section{\textcolor{black}{Further Considerations for Specifying CID Metric}} \label{sec:CIDmath}

\textcolor{black}{As noted previously, the specification of the CID metric should be tailored to the analysis and decision at hand.  Nonetheless, analysts can follow certain principles to assist development of the CID metrics, as we now describe.}

\textcolor{black}{We consider two parts to constructing a CID metric.  The first part concerns encoding the decision rule, as done, for example, in \eqref{decisionvote} and \eqref{eq:decthetahat}.   To establish actionable and realistic decision rules, analysts can work with domain experts or policy-makers.  In general, we expect them to be binary or ternary functions.}

\textcolor{black}{The second part concerns encoding the cost function that is incurred from inaccuracies in the estimate of interest. 
This specification is akin to specifying a cost (or utility) function, as done in decision analysis and in economics applications \citep{chung, cotomillan}.  In our applications, we use linear functions to represent costs of inaccuracies in analyses. In the Bread and Peace analysis, for example, this linear function implies that (i) analysts consider larger confidence interval overlaps to be more desirable than smaller confidence interval overlaps and (ii) analysts consider the marginal cost of the same decrease in the overlap to be constant.  Alternatively, analysts could specify a nonlinear function that expresses the costs of inaccuracies, for example, one that encodes a relatively flat cost when overlap exceeds 0.5 and drops sharply thereafter. We expect that, in many cases, analysts will want to enforce monotonicity when setting the cost function.}


\textcolor{black}{If analysts assign a constant cost function, 
effectively they just use the decision rule as the CID metric. For example, in the Bread and Peace analysis, the CID metric could be based purely on whether or not the prediction interval includes 0.5, irregardless of the confidence interval overlap.  In this case, the CID framework encourages analysts to consider whether the ultimate decision---not just the estimates---changes when assessing sensitivity to data imperfections. As noted previously, a 5\% change in a point estimate from a reference analysis can be interpreted differently depending on how close the point estimate is to the boundary of a decision change.} 

\textcolor{black}{Some decision rules also can be viewed through the lens of significance testing.  For example, in the North Carolina lead data analysis, with a constant cost function one could perform a significance test of the null hypothesis $\theta<0.20$ based on the multiple imputations for a given posited missingness distribution.  However, in other contexts significance testing may not be as natural to implement, for example, with  a three-option decision task like the one in the Bread and Peace CID analysis.}

\textcolor{black}{In economics applications and decision theory, analysts often use loss (utility)  functions to select from competing strategies.  The analyst selects the strategy that optimizes the loss function, according to some posited distribution of potential outcomes. Our sensitivity analysis context differs. First, if the analyst puts highest probability on a certain assumptions about the data imperfections, for example, a certain measurement error or missingness mechanism, it is rational for them to use that assumptions for the reference analysis.  The resulting CID metric would provide information on confidence in decisions for that reference analysis; optimization is not involved.  That said, given a probability distribution on the data imperfections, e.g., on the values of $t$ in our examples, it should be possible to compute an expected CID value.  This expectation can serve as a one-number summary of the confidence in decision metric for the reference analysis.}

 \section{Discussion} \label{sec:discussion}

Analysts using data to make policy decisions have to make choices about how to handle data imperfections.  In this article, we present and illustrate a framework that analysts can use to investigate their degree of confidence in their decisions under departures from the 


\noindent assumptions underlying their analysis choices.   We do so for the contexts of measurement error in the input to a predictive model and missingness due to selection effects.  The confidence in decision framework applies in other contexts as well.  For example, one could use the framework to assess the stability of conclusions to different modeling choices \citep{gelman:hennig, veridical}, to assess the effects on decisions of data perturbations introduced by a statistical agency as part of its  disclosure limitation practices \citep{yang:reiter:stability}, or to assess the sensitivity of decisions based on analyses of non-random samples \citep{groves2010total}. 

Our examples used analyses with limited numbers of variables. In these cases,  
departures from the assumptions used in the reference analysis can be encoded with single parameters. It may be possible to define univariate knobs in contexts with richer analyses.  For example, sensitivity analyses in causal inference of observational studies often posit a single unmeasured confounder. Starting from an analysis with no unmeasured confounding (the reference analysis), they determine how strong the confounder needs to be to make the estimated causal effect vanish (turning a knob until the decision changes).  \textcolor{black}{This is a CID analysis, which potentially could be enhanced with a metric that offers more nuance on the decisions, e.g., if different effect sizes result in different actions.} In some contexts, however, the analyst may seek to turn multiple knobs corresponding to multiple assumptions. 
Developing interpretable and practical methods for multivariate confidence in decision analysis is an important topic for future research.






\bibliographystyle{agsm}
\bibliography{bibliography.bib}

\end{document}